\documentclass[10pt,a4paper,twocolumn,superscriptaddress]{revtex4-2}

\usepackage{textcomp}
\usepackage{color}
\usepackage{amsmath}
\usepackage{mathrsfs}
\usepackage{graphicx}
\usepackage{subfig}
\usepackage{epsfig}
\usepackage[colorlinks]{hyperref}
\usepackage[percent]{overpic}

\usepackage{textcomp}
\usepackage[squaren,Gray]{SIunits}
\usepackage{tabularx}
\usepackage{float}
\usepackage{amssymb}
\usepackage[figurename=Fig.,labelfont=bf]{caption}
\usepackage{cleveref}
\usepackage{soul}
\usepackage{pgf}
\makeatletter

\newcommand{\customlabel}[2]{%
   \protected@write \@auxout {}{\string \newlabel {#1}{{#2}{\thepage}{#2}{#1}{}} }%
   \hypertarget{#1}{}
}
\makeatother

\begin{document}
\captionsetup[figure]{justification=raggedright}

\title{Dissipative solitons in parity-time symmetric laser cavities
} 


\author{Jes\'us Yelo-Sarri\'on}
\affiliation{OPERA-\textit{Photonics}, Universit\'e libre de Bruxelles (U.L.B.), 50~Avenue F. D. Roosevelt, CP 194/5, B-1050 Brussels, Belgium}
\affiliation{Departament de Física - IAC3, Universitat de les Illes Balears (U.I.B.), E-07122 Palma de Mallorca, Spain}

\author{Fran\c{c}ois Leo}
\affiliation{OPERA-\textit{Photonics}, Universit\'e libre de Bruxelles (U.L.B.), 50~Avenue F. D. Roosevelt, CP 194/5, B-1050 Brussels, Belgium}

\author{Simon-Pierre Gorza}
\email{simon.pierre.gorza@ulb.be}
\affiliation{OPERA-\textit{Photonics}, Universit\'e libre de Bruxelles (U.L.B.), 50~Avenue F. D. Roosevelt, CP 194/5, B-1050 Brussels, Belgium}

\begin{abstract}
\textbf{The generation of optically coherent ultrashort pulses by mode-locked lasers has revolutionized advancements in modern science and technology. These pulses often arise from the formation of dissipative solitons, which emerge due to a balance between energy excitation and dissipation. Harnessing the concept of parity-time (PT) symmetry to control this balance, we demonstrate a new type of laser dissipative solitons hosted in linearly coupled ring cavities. Our experiments are performed in a laser where the linear hybridized modes are in the PT-symmetric phase. We experimentally observe the formation of short pulses, stabilized by the selective breaking of the PT-symmetry by the Kerr nonlinearity. Our results unlock new possibilities for passive mode-locking by showing spontaneous pulse formation in PT-symmetric lasers without the need for additional intricate components such as saturable absorber or non-reciprocal and polarisation sensitive elements. }
\end{abstract}

\maketitle

In quantum mechanics, a system is said to be PT-symmetric when its associated Hamiltonian commutes with the parity(P)-time(T) operator. Although originally introduced in quantum field theory\,\cite{Bender1998}, the realisation that the concept of PT-symmetry could be applied to optics by engineering loss and gain has attracted a lot of attention\,\cite{Feng2017, Ozdemir2019, praveena_review_2023}. Beyond offering an ideal framework for exploring non-trivial fundamental effects and phase transitions in non-Hermitian Hamiltonians, PT-symmetry enables powerful mechanisms for controlling light flow, leading to remarkable phenomena such as coherent perfect absorption\,\cite{chong_coherent_2010, wong_lasing_2016}, non-reciprocal transmission\,\cite{chang_paritytime_2014, mao_-chip_2025}, and asymmetric reflectivity\,\cite{lin_unidirectional_2011, feng_experimental_2013}.

%
Dynamics in optical structures with a spatially antisymmetric gain and loss profile were initially investigated in Kerr nonlinear dual-core waveguides for optical switching\,\cite{chen_twin_1992} and nonlinear amplification
\,\cite{malomed_nonlinear-optical_1996}. These configurations were later recognized as PT-symmetric\,\cite{el-ganainy_theory_2007,guo_observation_2009, ramezani_unidirectional_2010, ruter_observation_2010}, sparking further exploration into arrays of coupled gain-loss waveguides\,\cite{musslimani_optical_2008} and to dispersive PT couplers\,\cite{alexeeva_optical_2012}.
%
Notably, stable solitons, appearing as beams or pulses, were theoretically predicted\,\cite{malomed_stable_1996, driben_stability_2011}, but have only been experimentally observed in PT-synthetic photonic lattices to date\,\cite{wimmer_observation_2015}.

The engineering of laser emission in PT-symmetric devices has also been widely explored, particularly in the vicinity of the phase transition occurring at the exceptional point (EP)\,\cite{liertzer_pump-induced_2012, Peng2014, hassan_nonlinear_2015, madiot_harnessing_2024}. 
This has paved the way for applications such as robust single-mode lasing\,\cite{hodaei_parity-time-symmetric_2014, feng_single-mode_2014, teimourpour_robustness_2017}, unidirectional lasing \,\cite{miao_orbital_2016}, and EP-based sensors with enhanced sensitivity\,\cite{hokmabadi_non-hermitian_2019, wiersig_review_2020, roy_nondissipative_2021}. 
However, most studies have focused on systems in the quasi-continuum regime, i.e. those supporting only a few modes, leaving the dynamics of multi-mode PT-lasers largely unexplored\,\cite{longhi_pt-symmetric_2016}, a gap that is especially relevant given their potential to advance ultrashort pulse generation and optical frequency combs.
%
%
%

Mode-locked (ML) lasers, by generating stable trains of ultrashort pulses\,\cite{haus_mode-locking_2000}, have revolutionized numerous fields, ranging from material processing to telecommunications or metrology\,\cite{sugioka_ultrafast_2014, keller_recent_2003, nahata_wideband_1996, fortier_20_2019}.   
%
%
Among various mode-locking techniques, passive mode-locking is particularly effective for generating high-quality short pulses. The two dominant approaches rely on either intrinsic saturable absorbers (SA) --- such as semiconductor saturable absorbers\,\cite{keller_semiconductor_1996} or graphene\,\cite{bao_atomic-layer_2009}--- or additive pulse mode-locking (APM), initially implemented using an auxiliary nonlinear cavity\,\cite{haus_structures_1991,keller_coupled-cavity_1990, belanger_coupled-cavity_1991} and later refined into nonlinear polarisation rotation\,\cite{tamura_self-starting_1992} and nonlinear optical loops\,\cite{duling_all-fiber_1991, fermann_nonlinear_1990}.
%
%
Interestingly, stable pulse solutions have been theoretically predicted in two linearly coupled Ginzburg-Landau equations, one with gain and the other loss (see\,\cite{malomed_solitary_2007} and references therein), a configuration inherently PT-symmetric. This suggests that PT symmetry could offer new strategies for mode-locking, particularly in systems where conventional APM techniques are challenging, such as integrated solid-state and diode lasers\,\cite{imamura_exceptional_2024}. Indeed, integrated passive ML sources currently rely on intrinsic SA\,\cite{van_gasse_recent_2019, okawachi_chip-scale_2023}, which is known to limit noise and timing jitter performances\,\cite{hansel_all_2017}. Alternative mode-locking strategies are thus highly desirable.
%
%

Here, we demonstrate how the modal properties of PT-symmetric active cavities can be leveraged to achieve passive mode-locking operation. Specifically, we experimentally explore the formation of pulses, as dissipative solitons, in a ring laser cavity coupled to a passive one. 



%
\begin{figure*} 
\customlabel{fig:1}{1}
\customlabel{fig:1a}{1a}
\customlabel{fig:1b}{1b}
\customlabel{fig:1c}{1c}
\customlabel{fig:1d}{1d}

    \centering
    \includegraphics[width=\textwidth]{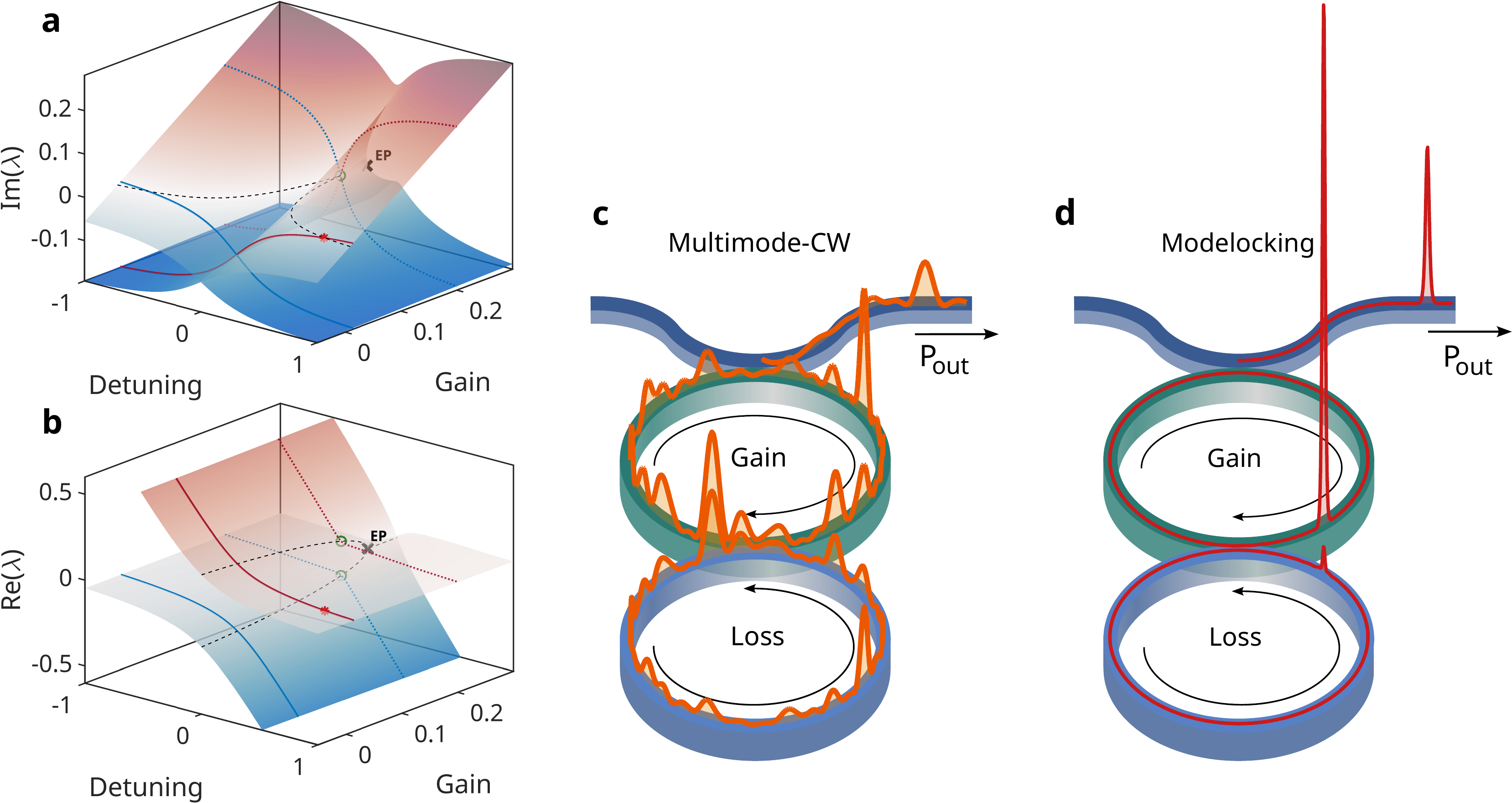}
    \caption{\textbf{Principle of mode-locking in PT-symmetric lasers.} 
    \textbf{a, b}, Eigenvalue surfaces in the (detuning, gain)- parameter space for constant loss $\alpha = 0.2$, below the coupling $\kappa = 0.236$. The complex eigenfrequencies of the supermodes are $\omega_{1,2}=\lambda_{1,2}/t_R$, with $t_R$ the cavity round-trip time. The supermodes experience a net positive gain when $\mathrm{Im}(\lambda)>0$. At zero detuning, the PT symmetry is broken beyond the gain value ($g_\mathrm{EP} = 0.272$) of the exceptional point (EP). Dashed curve in (a): $\mathrm{Im}(\lambda)=0$. Dashed curves in (b): solutions at zero detuning.  
    \textbf{c}, For phase-matched cavities ($\delta_0=0$), the laser remains in the unbroken phase, corresponding to quasi-CW multimode emission, providing that the nonlinear detuning ($\delta_\mathrm{NL}$) is also null, i.e. for $P_\mathcal{G} = P_\mathcal{L}$. The laser reaches equilibrium through gain saturation at the exact PT condition $g=\alpha<\kappa$ (green circles in a-b).
    \textbf{d}, A detuning between the coupled resonators breaks the symmetry, leading to the general stationary condition for the intra-cavity energies ($E_\mathcal{G,L}$), $g E_\mathcal{G} = \alpha E_\mathcal{L}$ (red star in a-b). This can be exploited to stabilize soliton mode-locked pulses in the broken phase through the Kerr detuning induced by the power imbalance between the resonators. The gain saturation ensures $\mathrm{Im}(\lambda)<0$ for the background modes in the PT-symmetric phase.      
    }
    \label{fig:principle_PT-ML}
\end{figure*}

\textbf{Concept.} Let us consider an arrangement of two identical ring resonators with equal resonant frequencies, but with one ($\mathcal{R_G}$) exhibiting a gain $g$ and the other ($\mathcal{R_L}$) some loss $\alpha$ (see Fig.~\ref{fig:1}). These resonators are linearly coupled with a coupling strength $\kappa$.
Under the condition that the gain-loss distribution is exactly balanced ($g=\alpha)$, the system can be considered PT-symmetric\,\cite{hodaei_parity-time-symmetric_2014, Peng2014}. Remarkably, for $\kappa$ larger than the gain, the supermodes arising from the coupling are evenly spread on both resonators (unbroken or PT-symmetric phase) and remain neutral. Conversely, beyond the threshold $g>\kappa$, a phase transition to the broken PT-symmetry regime occurs, with the growing and decaying modes predominantly confined to the gain and lossy resonators, respectively.
%
The eigenfrequencies associated with the supermodes, $\omega_{1,2}$, 
are real in the PT-symmetric phase, but complex in the general case as shown in Fig.~\ref{fig:1}a-b, with the real (imaginary) part describing the mode frequency shift (amplification/attenuation). 

When the stringent requirement of loss-gain balance is not met, the notion of PT-symmetry can be generalized by changing the system reference point from zero (exact PT) to the gain-loss average $\chi = (g-\alpha)/2$ (hidden-PT), leading to a global exponential amplification  ($\chi>0$) or decay ($\chi<0$)\,\cite{Ozdemir2019}. For fixed loss and coupling, increasing the gain reveals the transition between the split resonance (PT-symmetric) and split dissipation (PT-broken) regimes, which occurs at $g_\mathrm{EP}=2\kappa-\alpha$ (see Fig.~\ref{fig:1a} at zero detuning). 
Nonetheless, for $\alpha<\kappa$, the gain saturation could prevent the transition to the broken phase. The condition for the total energy to be stationary is $gE_\mathcal{G} = \alpha E_\mathcal{L}$\,\cite{alexeeva_optical_2012}, with $E_\mathcal{G,L}$, the energy in each cavity. For an equal energy, this condition gives $\alpha=g=g_0/(1+E_\mathcal{G}/E_{s})$, with $g_0$ the unsaturated gain and $E_s$ the saturation energy of the gain medium (green circles in Fig.~\ref{fig:1a}-\ref{fig:1b}). Lasing on many pairs of modes is expected for coupled resonators supporting multiple longitudinal modes and an active medium with an inhomogeneously broaden gain profile $g(\omega)$ (Fig.~\ref{fig:1c}), unless a special design that exploits transition at the exceptional point is implemented\,\cite{hodaei_parity-time-symmetric_2014}.   

The PT-symmetry can also be broken by shifting the relative detuning between the resonators away from zero. In the linear propagation regime, the detuning is $\delta_0 = (\omega_\mathcal{G}-\omega_\mathcal{L})\times t_R$, with $\omega_\mathcal{G,L}$ the resonant frequencies of the uncoupled resonators and $t_R$ the cavity round-trip time. 
%
%
The supermode confined in the active resonator then experiences more gain (or less dissipation) than the one in the unbroken phase at $\delta_0=0$ (red star in Fig.~\ref{fig:1a}).       
The key idea underlying mode-locking in PT-symmetric dispersive resonators is that the Kerr effect can selectively break the PT symmetry\,\cite{yelo-sarrionKerrCavitySolitons2023, imamura_exceptional_2024}. 
The imbalance between the powers $P_\mathcal{G,L}$ in the cavities results in an additional detuning $\delta_\mathrm{NL} = \gamma(P_\mathcal{G}-P_\mathcal{L})L$, with $\gamma$ the nonlinear Kerr coefficient and $L$ the cavity length, lowering the loss for strong pulses in the active ring coupled to weak pulses in the passive one. 
We thus foresee the formation of short pulses in the anomalous dispersion regime (Fig.~\ref{fig:1d}), as the Kerr effect stabilizes nonlinear localized structures in the form of fundamental solitons\,\cite{atai_stability_1996, malomed_solitary_2007}. 
%
%

\begin{figure*} 
\customlabel{fig:2a}{2a}
\customlabel{fig:2b}{2b}
\customlabel{fig:2c}{2c}
\customlabel{fig:2d}{2d}
\customlabel{fig:2e}{2e}
\customlabel{fig:2f}{2f}
    \centering
    \includegraphics[width=\textwidth]{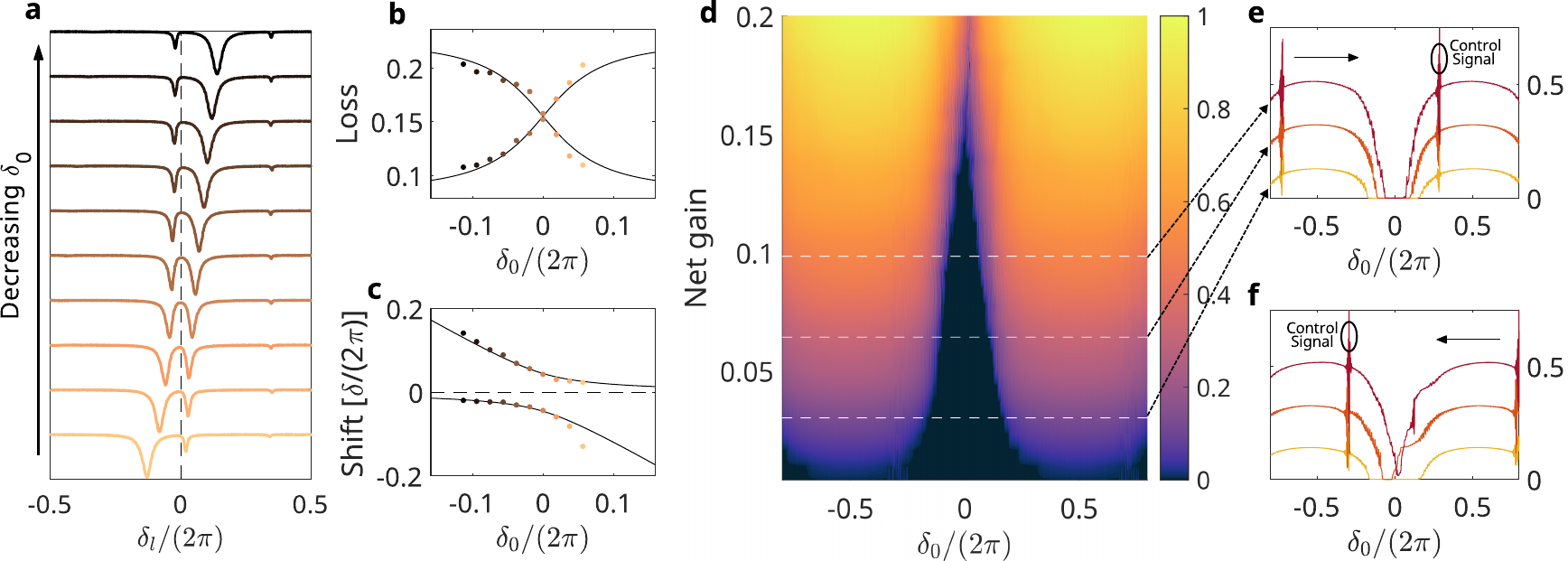}
    \caption{ 
    \textbf{Effect of the detuning and the gain.} \textbf{a}, Transmission spectra measured in the through port of $\mathcal{R_G}$ with an external laser by sweeping its detuning ($\delta_l$). The net gain in the active cavity is negative to avoid lasing. 
    \textbf{b}, Modal loss of the two supermodes extracted from the transmission curves in (a). The solid line shows the theoretical curve ($\tilde{g} =-0.45$).
    \textbf{c}, Same but for the resonance shifts. These curves (which correspond to the imaginary and real parts of the mode eigenfrequencies, respectively) illustrate how the detuning $\delta_0$ controls the modal loss and the resonant frequencies. 
    \textbf{d}, Normalized laser signal for increasing values of the net gain, measured for forward detuning scans.
    \textbf{e}, Laser signal for three particular gain values. The laser switches off in the vicinity of the zero detuning where the modal losses are the highest. \textbf{f}, Backward scans uncover a bistable behaviour. It is associated with the spontaneous generation of dissipative solitons. The narrow peaks in (e-f) come from the control signal needed to stabilize the detuning. This peak was numerically filtered in (d) for clarity.       
    }
    \label{fig:scan}
\end{figure*}

\textbf{Results.} The experimental demonstration 
is performed with fibre ring resonators. The setup is schematically represented in Fig.~\ref{fig:principle_PT-ML}. It mainly consists of an active cavity, about 18\,m long, coupled to a passive cavity of identical length ($\kappa=\sqrt{0.1}$, $\alpha=0.17$).  
%
%
To observe the emergence of mode-locking, the system is parametrically controlled via the linear detuning $\delta_0$, and an additional loss $\alpha'$ induced in $\mathcal{R}_\mathcal{G}$ to tune the net gain $\tilde{g} = g-\alpha'$.      

In the first set of experiments, we begin with a negative net gain to prevent lasing and we measure the resonances of the coupled resonators. This is performed in the through port of $\mathcal{R}_\mathcal{G}$ by using an external probe laser whose frequency can be tuned to sweep its detuning $\delta_l$. At exact resonance matching ($\delta_0 = 0$, corresponding to unbroken PT symmetry), both supermodes experience equal losses, and their resonances are symmetrically split by $\kappa$. As seen in Figs.~\ref{fig:2a}-\ref{fig:2c}, when the detuning is moved away from $\delta_0=0$, one mode becomes more lossy while its resonance shifts away from $\delta_l = 0$. In contrast, the resonance of the other mode moves toward zero, and exhibits reduced loss.


We then switch off the probe laser and gradually increase the net gain while scanning the detuning. In Figs.~\ref{fig:2d}-\ref{fig:2e}, we show the power measured at the active cavity output port during a forward sweep of the detuning. Depending on the level of gain, the system either remains below the lasing threshold, stays above it, or crosses the threshold twice, with the trivial state observed around the exact matching condition. 
These first experiments confirm that the detuning parameter governs the dissipation of the supermodes, as predicted by their complex eigen-frequencies.


%
Scanning the detuning in the reverse direction reveals the existence of a bistable behaviour. As the net gain increases, the power curve with $\delta_0$ becomes asymmetric (Fig.~\ref{fig:2f}). Noticeably, lasing occurs for positive detunings in the region where the laser was previously below the threshold. This bistable behaviour strongly depends on mutual cavity synchronization, highlighting the critical role of resonance splitting across all longitudinal modes. The observed dynamics suggest the involvement of multiple interacting modes, which enable the formation of localized structures, as seen in the numerical simulations of forward and backward scans shown in Supplementary Fig.2.

\begin{figure*}[]
\customlabel{fig:3a}{3a}
\customlabel{fig:3b}{3b}
\customlabel{fig:3c}{3c}
\customlabel{fig:3d}{3d}
\customlabel{fig:3e}{3e}
\customlabel{fig:3f}{3f}
\customlabel{fig:3g}{3g}
\customlabel{fig:3h}{3h}
\centering
\includegraphics[width=\textwidth]{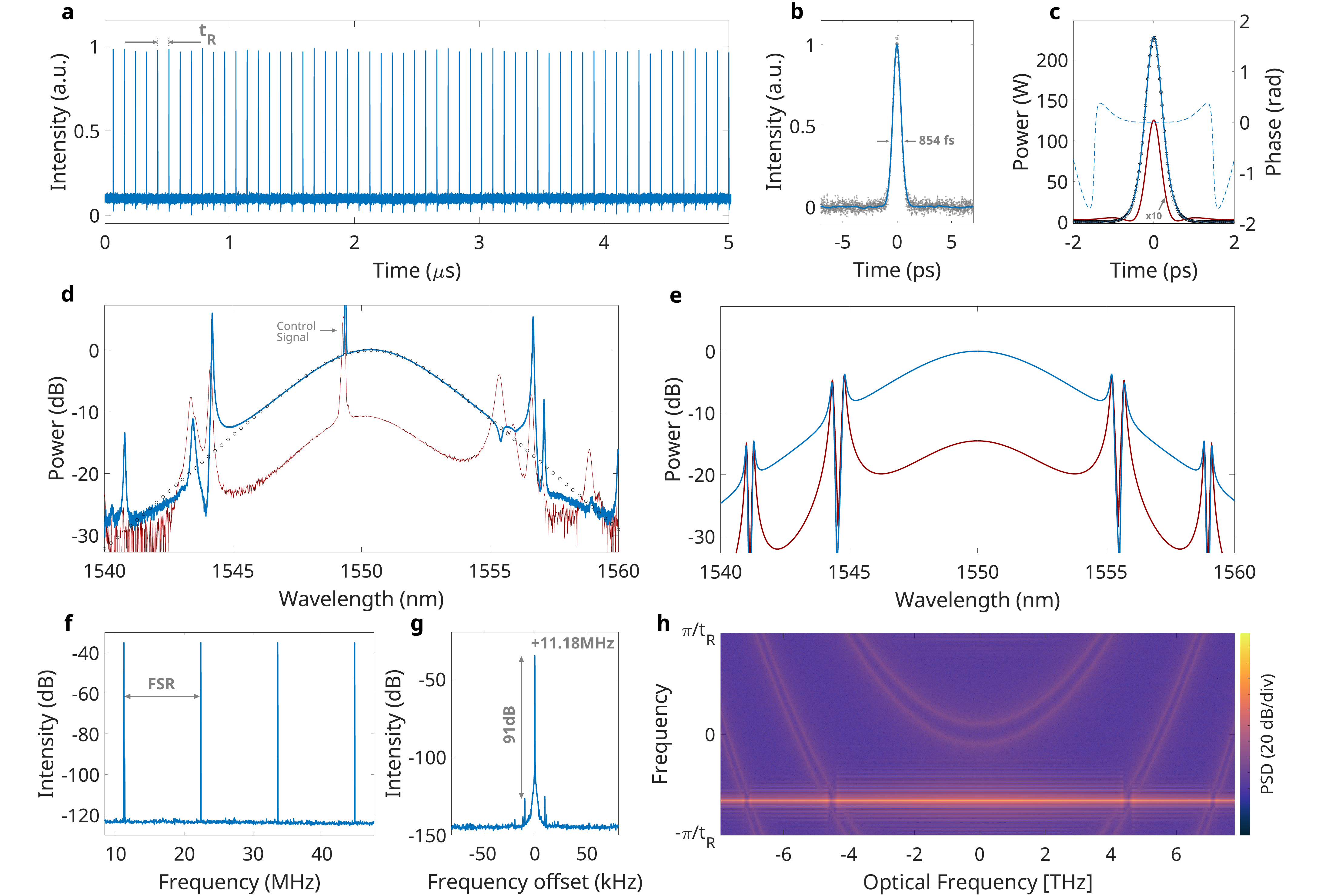}
\caption{
\textbf{Mode-locking characterisation}.
\textbf{a}, Oscilloscope recording showing that one short pulse leaves the active cavity every round-trip time. 
\textbf{b}, Experimental autocorrelation trace of the output pulses.
\textbf{c}, Theoretical profiles of the pulses in the active (blue) and passive (red, $\times10$) resonators. The solid lines represent the powers and the dashed line the phase, computed from the laser lumped model (see Methods), whereas the circles correspond to a fundamental soliton.
\textbf{d}, Experimental spectra at the output of the active (blue) and passive (red) cavities. The circles represent the spectrum of a 605\,fs hyperbolic secant pulse. The peaks at 1549.4\,nm come from the control signals.
\textbf{e}, Theoretical spectra of the pulses plotted in (c).
\textbf{f}, An experimental radiofrequency (RF) spectrum of the pulse train leaving the active cavity, showing beat notes separated by the FSR. 
\textbf{g}, High-resolution RF spectrum around the fundamental beatnote.
\textbf{h}, Theoretical 2D power spectral density (PSD) map, 
obtained from the 2D Fourier transform along the fast and slow time coordinates of the mode-locked pulse for $\delta_0=0$. The folded split parabolas reveal the effective dispersion of the two families of supermodes. The positions of the Kelly sideband pairs in the spectra are defined by the intersection points of the split parabolas with the PSD of the dissipative soliton. 
}
\label{fig:3}
\end{figure*}

To characterize the new lasing state, we stabilize the detuning near the exact matching condition and increase the net gain above the lasing threshold. Within a small range of delays between the two resonators, the output optical spectra transitions from narrowband to broadband. In the latter case, the temporal signal detected by a fast photodiode and observed on a 10\,GHz oscilloscope reveals the emergence of short pulses. 
Yet, because of the average power required for gain saturation, irregular patterns of multiple pulses per round-trip time were observed.
%
We thus reduce the net gain until a single pulse leaves $\mathcal{R}_\mathcal{G}$ every round-trip, and record the photodiode signal, the autocorrelation trace, as well as the optical spectra. Our results, reported in Fig.~\ref{fig:3}, demonstrate that a single 600\,fs pulse around 1550\,nm circulates in the active cavity with a power imbalance between the two resonators of about $4.5$\,dB. Moreover, the overall shape of the recorded spectrum in the active cavity and the autocorrelation trace agree well with a hyperbolic secant pulse profile, consistent with the expectation of soliton formation. 


Numerical simulations provide insight into the pulse dynamics.
The simulated spectra with the parameters of our cavity globally reproduce the characteristic features seen in the experiment (see Fig.~\ref{fig:3e}). The analysis of the corresponding temporal profile in $\mathcal{R}_\mathcal{G}$ shows that, in the steady state, a 230\,W peak fundamental soliton is generated in the active cavity. This pulse is coupled to a phase-locked low amplitude pulse in $\mathcal{R}_\mathcal{L}$, supporting the mode-locking principle whereby the generated pulses are in the PT broken phase due to the nonlinear detuning. 
Conversely, the linear modes remain in the unbroken phase with $\mathrm{Im}(\lambda$) negative (see plain curves in Fig.~\ref{fig:1a}), thereby stabilizing the trivial background solution.
%
We note that in our experiment, the strong coupling ($\kappa > \alpha$) sets the PT-laser below the exceptional point ($\tilde{g}<g_\mathrm{EP}$). Stable mode-locking is also expected for $\tilde{g}>g_\mathrm{EP}$\,\cite{imamura_exceptional_2024}, but in that regime, the modal loss contrast is smaller.

The experimental spectra recorded in the two resonators exhibit narrowband peaks in the wings, in addition to the central peak associated with the control signal. These are Kelly sidebands. They are commonly encountered in soliton mode-locked lasers and arise from the periodic perturbations of the pulse over the cavity round-trip\,\cite{kelly_characteristic_1992, smith_sideband_1992, dennis_experimental_1994, haus_mode-locking_2000}. Notably, here the Kelly sidebands do not appear as isolated peaks. Instead, they have a double-peak structure as seen in Fig.~\ref{fig:3d} around 1544\,nm and in the numerical simulation in Fig.~\ref{fig:3e}. The origin of the peak pairs lies in the hybridisation of the modes and the resulting resonance splitting (see Fig.~\ref{fig:2a} and Fig.~\ref{fig:1b} for $\tilde{g}<g_\mathrm{EP}$). Fig.~\ref{fig:3h} displays the 2D Fourier transform of the simulated intra-cavity field in $\mathcal{R}_\mathcal{L}$, recorded over numerous round-trips once the pulse has stabilized. As can be seen, the small noise from the spontaneous emission in the amplifier reveals the effective dispersion relation\,\cite{leisman_effective_2019, anderson_dissipative_2023}. The splitting forms two distinct optical frequency bands: one for the symmetric modes and one for the antisymmetric modes\,\cite{tikan_emergent_2021}.
They repeat every $2\pi/t_R$ because of the round-trip periodic nature of the perturbations. The 2D power spectral density map, where the mode-locked pulse appears as a straight line, uncovers that the pairs of Kelly sidebands stem from the intersection between the pulse and the split linear folded bands. As a result, the resonant photon transfer to each of the two mode families leads to the formation of multiple pairs of Kelly sidebands within the soliton spectrum. The double peak structures seen in the spectra are thus the signature that the linear cavity modes are in the unbroken phase. It should be noted that the extra peaks around 1555.5\,nm seen in the experiment, mainly in $\mathcal{R}_\mathcal{L}$ and which have no equivalent on the other side of the spectra, are not explained by the simulations. They probably come from unwanted coupling effects between not perfectly aligned polarisation modes in the two fibre cavities.

The stability of the pulse train is characterized by directing the signal leaving $\mathcal{R_G}$ into a photodiode connected to a radio-frequency (RF) analyzer. The spectra, as seen in Figures~\ref{fig:3f}-\ref{fig:3g}, show narrow peaks spaced by the free spectral range of the individual cavities. The high signal-to-noise ratio and the absence of significant satellite peaks confirm the generation of a stable output pulse train, a key result for practical applications. We note that breathing dynamics of the mode-locked solution in the broken phase have been observed in simulations for other cavity parameters (see Supplementary Fig.3). These dynamics are likely linked to the localized oscillatory states predicted in PT-symmetric nonlinear couplers\,\cite{alexeeva_optical_2012} and coherently driven PT-dimers \,\cite{Milian2018}, as well as to the self-pulsing behaviour of the out-of-equilibrium steady-state in coherently driven coupled resonators\,\cite{carlon_zambon_parametric_2020,yelo-sarrion_self-pulsing_2021}.

\begin{figure}[]

\customlabel{fig:4a}{4a}
\customlabel{fig:4b}{4b}
\customlabel{fig:4c}{4c}
    \centering
\includegraphics[width=\columnwidth]{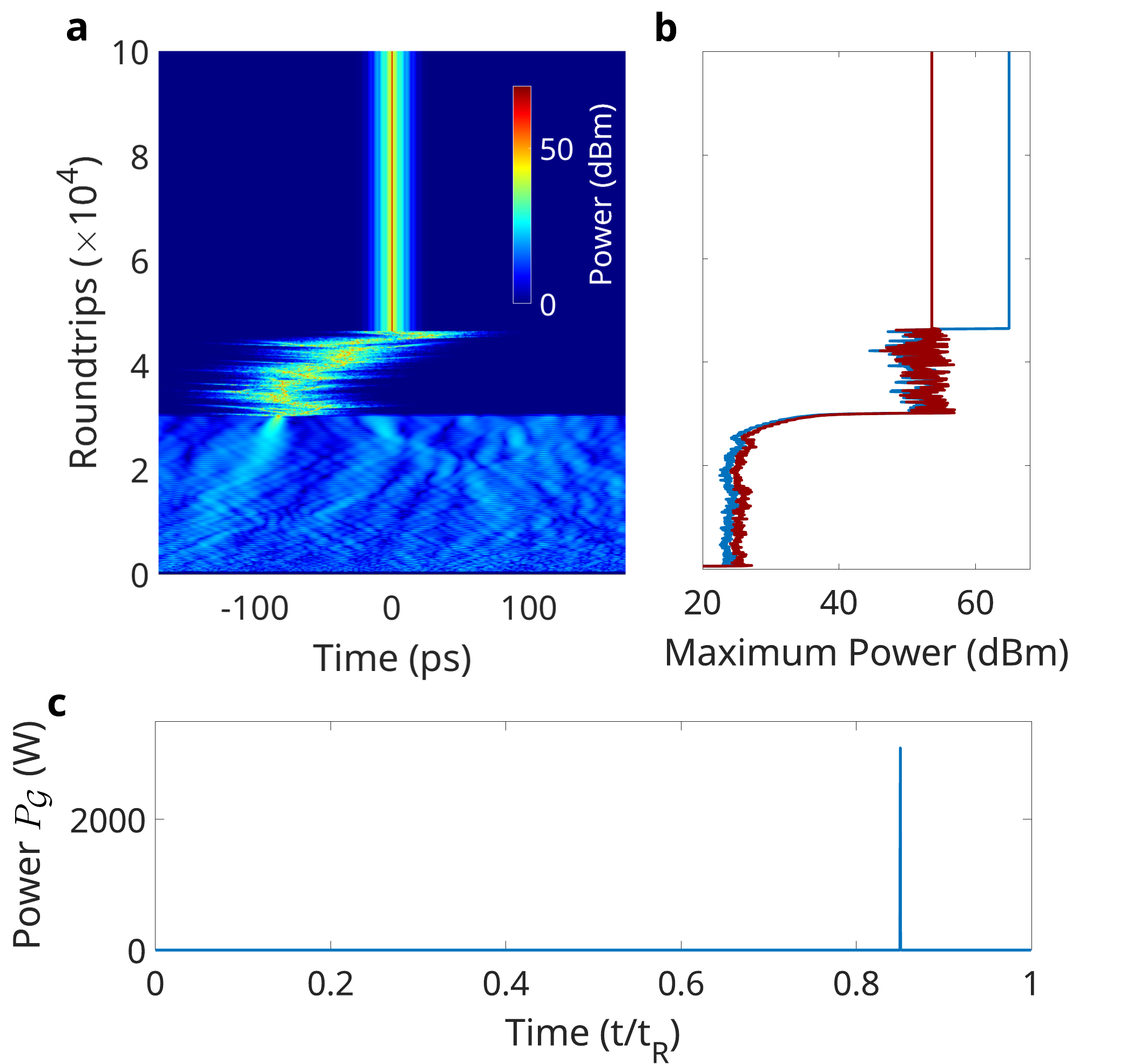}
\caption{Simulation of the self-starting mode-locking dynamics. 
\textbf{a}, Temporal dynamics of spontaneous mode-locking from noise, zoomed around the emerging pulse in the active cavity. 
\textbf{b} The evolution of the maximum powers shows that the powers are initially nearly evenly distributed between both resonators ($P_\mathcal{G}^\mathrm{max}$, blue and $P_\mathcal{L}^\mathrm{max}$, red). The unstable regime evolves toward a stable dissipative soliton in the PT-broken phase.  
\textbf{c} Intra-cavity power after 10$^5$ round-trips, highlighting the generation of a single pulse in the active cavity. The round-trip time is $t_R=8.9$\,ns. }
\label{fig:4_starting_dynamics}
\end{figure}

At zero detuning and below the EP ($\tilde{g}<g_\mathrm{EP}$), the laser field initially grows on the (linear) supermodes that are equally spread on both resonators. Yet, it is the power imbalance between the cavities that stabilizes the pulsed solution. This raises the question of the self-starting capability of mode-locking in PT-symmetric lasers. To investigate the starting dynamics and the spontaneous pulse formation, we numerically simulate the evolution of the intra-cavity fields (we consider a round-trip time 10$\times$ shorter than in the experiment for numerical reasons).  
Figure~\ref{fig:4_starting_dynamics} shows that in the first stage, the intra-cavity powers quickly increase to reach a first plateau. 
In that state, the laser operates in a multimode, incoherent regime, characterized by multiple temporal peaks filling both cavities and an almost equal power distribution in the resonators. 
In the second stage between $\sim [2.5-4.5]\times10^4$ round-trips, a single localized structure takes over and becomes predominant over continuous-waves. The pulse, for which the powers in the two cavities are still comparable, is highly unstable with large amplitude and position oscillations. This unstable regime persists until the system finally reaches the PT-broken phase at round-trip $4.65\times10^4$. Beyond, a single stable 140\,fs pulse emerges.                    

\textbf{Discussion.} 
We have demonstrated stable mode-locking operation in a parity-time symmetric ring pair. We have shown that by harnessing the properties of the hybridized modes, nonlinear coupled cavities may host stable dissipative solitons. Recently, novel architectures utilizing tailored two-cavity configurations have been explored for making comb sources\,\cite{bao_laser_2019,letsou_lasing_2024}. 
Here, the mode hybridisation and the nonlinear detuning acquired in the PT-broken phase play a central role in stabilizing the mode-locked pulse.
Unlike reciprocal Sagnac loops, such as in Figure-of-Eight or Figure-of-Nine laser cavities\,\cite{hansel_all_2017}, PT-symmetric lasers do not require a nonreciprocal phase bias. Furthermore, the inherent non-zero slope of the modal loss with respect to the power imbalance, below the EP, enhances the self-starting capability of pulse formation.    
In the experiment, we observed the spontaneous emergence of dissipative solitons either at zero detuning with a gain above the CW threshold, or with a slightly lower gain by scanning the detuning between the resonators.  

Our demonstration of mode-locking in PT-symmetric lasers was made with two coupled fibre ring resonators and erbium gain. However, this pulse formation principle is universal, and we anticipate its demonstration on other platforms, such as in integrated lasers. 
With recent advances in solid-state integrated gain media\,\cite{wang_photonic-circuit-integrated_2023, yang_titaniumsapphire--insulator_2024, liu_photonic_2022} and hybrid integrated diode lasers\,\cite{van_gasse_recent_2019}, there is a growing need for self-starting, robust, and efficient mode-locking schemes\,\cite{shtyrkova_integrated_2019} with low intrinsic noise to drive the next generation of integrated pulse sources and optical frequency combs\,\cite{hermans_-chip_2022}.
Our results provide a path to develop the next generation of integrated passively mode-locked pulse sources and optical frequency combs.
Such an approach overcomes challenges associated with standard techniques, which are difficult to adapt due to the need for a nonreciprocal phase bias\,\cite{lau_development_2024}.

\textbf{Acknowledgments.}
We are grateful to C. Mas Arab\'{i} and N. Englebert for fruitful discussions. 
This work was supported by the Fonds de la Recherche Scientifique” (F.R.S.-FNRS) and the FWO under the Excellence of Science (EOS, 40007560) program, from the F.R.S.-FNRS (PDR.T.0104.19, CDR, J.0079.23), and by the European Research Council (ERC) under the Horizon Europe research and innovation program (Grant Agreement No. 101113343). F.L. acknowledges the support of the F.R.S.-FNRS. J.Y.S. acknowledges the support from project KEFIR/AEI/10.13039/501100011033/FEDER, UE.

\textbf{Author Contributions}
S-P.G conceived the experiment, and J.Y.S performed the experiments, supervised by F.L and S-P.G. J.Y.S and S-P.G performed the simulations. S-P.G supervised the overall project and wrote the manuscript. All authors discussed the results and contributed to the final manuscript.

\textbf{Competing Interests}
The authors declare no competing interests.

\bibliography{PT-ML_biblio}

\vspace{0.5cm}

\textbf{Methods}
{\footnotesize

\textbf{Experimental setup} 
A schematic of the full experimental setup is shown in the Supplementary Fig.1. The two resonators, mainly made of standard silica telecommunication fibre, are coupled by a 90/10 coupler ($\kappa = 0.316$). They both have a length of approximately 18\,m, leading to a cavity free spectral range (FSR) of 11.18 MHz. The gain in the active cavity, $\mathcal{R}_\mathcal{G}$, is provided by a section of $50$\,cm of erbium-doped fibre (EDF, Liekki\textsuperscript{\tiny{\circledR}} ER16-8/125). It is backward pumped at 1,480\,nm through a first wavelength division multiplexer (WDM), and the non-absorbed pump power is extracted by a second WDM. All experiments were carried out at a fixed pump power of 2\,W. A polarisation insensitive isolator, located directly after the second WDM, ensures lasing in only one direction. 
A fibred variable optical attenuator (VOA) in the active cavity controls the net gain $\tilde{g}$ and thus the gain-loss balance between the coupled resonators.
$\mathcal{R}_\mathcal{G}$ includes a 99/1 coupler, and $\mathcal{R}_\mathcal{L}$ includes both a 99/1 and a 95/5 coupler to extract part of the intracavity power and for stabilisation purposes. The measured finesse of $\mathcal{R}_\mathcal{L}$ is $\mathcal{F} = 14$ (Q factor of 2.4$\times10^8$), which corresponds to $\alpha_\mathcal{L} = 0.17$.  
Finally, each cavity includes a piezoelectric fibre stretcher and a polarisation controller. The stretchers were used to tune and stabilize the cavity detunings independently. The polarisation controllers set the polarisation states in the two cavities to avoid cross-polarisation coupling. 

To coarsely synchronize the two resonators, we measured and adjusted the length of $\mathcal{R}_\mathcal{L,G}$ through their respective FSR. The optimal synchronisation was then obtained by fine-tuning the length of $\mathcal{R}_\mathcal{G}$ with a variable optical delay line by maximizing the bistability region when scanning the detuning [see Figs.\ref{fig:scan}(e-f)]. 
The relative detuning between the resonators ($\delta_0$) was controlled by independently stabilizing the two fibre cavities with the Pound-Drever-Hall (PDH) method. 
First, the length of $\mathcal{R}_\mathcal{L}$ was interferometrically stabilized to a sub-100 Hz linewidth continuous-wave (CW) Koheras Adjustik laser. This was achieved by sending part of the CW laser output to a frequency-shifter (FS1) followed by a phase modulator driven at half the cavity FSR.
This signal was fed to the 99/1 tap coupler in the passive cavity, in the direction opposite to lasing to probe the resonance of $\mathcal{R}_\mathcal{L}$, decoupled from $\mathcal{R}_\mathcal{G}$, thanks to the isolator.     
%
%
The PDH error signal was then sent to a proportional-integral-derivative (PID) controller that drives the intracavity piezoelectric stretcher through a high-voltage amplifier.
The remaining part of the CW laser beam was sent to another frequency shifter (FS2) and phase modulator, and fed to the same 99/1 tap coupler but in the forward direction. It thus resonates within the coupled cavities, providing a PDH feedback signal to lock $\delta_0$ once $\mathcal{R}_\mathcal{L}$ is stabilized. The detuning $\delta_0$ was then tuned by changing the driving frequency of FS2.    
The measurements reported in Fig.~\ref{fig:scan}(a) were obtained by locking both $\delta_\mathcal{L}$ and $\delta_0$ and by scanning the frequency of an external continuous wave laser. In Figs.\,\ref{fig:scan}(d-f), we recorded the laser output power with $\delta_\mathcal{L}$ locked and while scanning $\delta_\mathcal{G}$ by applying a voltage ramp to the piezoelectric stretcher in $\mathcal{R}_\mathcal{G}$. In experiments reported in Fig.~\ref{fig:3} both $\mathcal{R}_\mathcal{L}$ and $\mathcal{R}_\mathcal{G}$ were stabilized to set $\delta_0\approx0$. We note that the width of the autocorrelation trace is 10$\%$ shorter than predicted from the experimental spectrum due to a small nonlinear temporal focusing experienced in the amplifier used before the autocorrelator.

\textbf{Numerical results} 
The simulation results shown in Figs.\,\ref{fig:3},\ref{fig:4_starting_dynamics} were obtained by using a lumped model of the coupled cavities. The propagation in the fibre resonators was simulated by integrating the nonlinear Schrödinger (NLS) equation. We then applied the transmission coefficients $\sqrt{T_\mathcal{L}}, \sqrt{T_\mathcal{G}}$ to the fields to account for the losses in both cavities. The saturated amplification in the gain medium was considered by applying the amplification factor $\sqrt{G}$, with $G = \exp{\left[g_0/(1+\bar{P}_\mathcal{G}/P_\mathrm{sat})\right]}$, where $\bar{P}_\mathcal{G}$ is the average power in the active cavity, $P_\mathrm{sat}=220$\,mW is the saturation power of the amplifier, and $g_0$ is the small signal gain ($g_0=1$ in Fig.~\ref{fig:3} and 1.3 in Fig.~\ref{fig:4_starting_dynamics}). We included the amplifier noise by adding in the spectral domain the equivalent of one virtual photon per modes. Finally, the boundary condition on the coupler and the detuning $\delta_0$ were applied to the intra-cavity fields. In the simulation reported in Fig.~\ref{fig:4_starting_dynamics}, spectral Gaussian filtering (5.6\,THz bandwidth) was also applied. The simulation parameters are provided in the Supplementary Section 2.        }     

\end{document}


\title{Supplementary Information for\\
Dissipative solitons in parity-time symmetric laser cavities
} 

\author{Jes\'us Yelo-Sarri\'on}
\affiliation{OPERA-\textit{Photonics}, Universit\'e libre de Bruxelles (U.L.B.), 50~Avenue F. D. Roosevelt, CP 194/5, B-1050 Brussels, Belgium}
\affiliation{ Departament de Física - IAC3, Universitat de les Illes Balears (U.I.B.), E-07122 Palma de Mallorca, Spain}

\author{Fran\c{c}ois Leo}
\affiliation{OPERA-\textit{Photonics}, Universit\'e libre de Bruxelles (U.L.B.), 50~Avenue F. D. Roosevelt, CP 194/5, B-1050 Brussels, Belgium}

\author{Simon-Pierre Gorza}
\affiliation{OPERA-\textit{Photonics}, Universit\'e libre de Bruxelles (U.L.B.), 50~Avenue F. D. Roosevelt, CP 194/5, B-1050 Brussels, Belgium}


\maketitle


\section{Experimental setup}

\begin{figure}[h!]
    \centering
     \includegraphics[width=\textwidth]{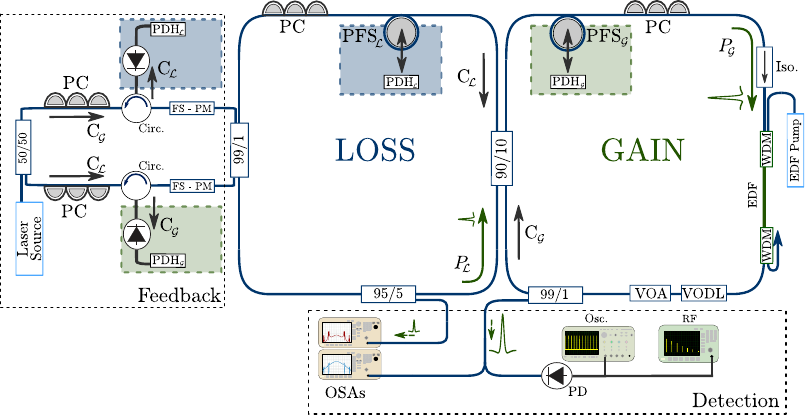}

    \caption{\textbf{Experimental setup.} 
    PC, polarisation controller, Circ., optical circulator, Iso., optical isolator, WDM, 1480/1550 wavelength division multiplexer, EDF Pump, 1480\,nm laser for pumping the erbium-doped fibre (EDF), VOA, variable optical attenuator, VODL, variable optical delay line, PFS, piezoelectric fibre stretcher, PD, photodiode, OSA, optical spectrum analyser, Osc., Oscilloscope, RF, radio-frequency analyser. 99/1, 95/5 and 90/10 are fibre couplers.      
    FS-PM, frequency shifter followed by a phase modulator for generating the control signals $C_\mathcal{L,G}$ used to stabilize the lossy cavity detuning ($\delta_\mathcal{L}$) and the relative detuning $\delta_0$, respectively. A Pound-Drever-Hall scheme was implemented for their stabilisation.    
    $P_\mathcal{L,G}$, laser power in the active ($\mathcal{G}$) and passive ($\mathcal{L}$) cavities.
    }
    \label{fig:exp-setup}
\end{figure}

The two ring resonators, mainly made of standard silica telecommunication fibre, have a length of 18.23\,m.
They are coupled by a 90/10 coupler. The other couplers are used to extract part of the intra-cavity laser powers and for stabilisation purposes.
The cavities are phase stabilized by means of piezoelectric fibre stretchers.
The gain in the active cavity is provided by a section of $\approx50$\,cm of erbium-doped fibre (EDF, Liekki\textsuperscript{\tiny{\circledR}} ER16-8/125), pumped by a 2\,W, 1480\,nm laser. 
%
%
The variable optical attenuator enable to tune the gain in the active cavity.
The polarisation controllers set the polarisation states to avoid cross-polarisation coupling. We note that the isolator is polarisation insensitive 
and that the mode-locking regime is highly sensitive to the synchronisation between the two cavities.  

\newpage
\section{Lumped-element model parameters}
The parameters of the lumped-element model used in the numerical simulations are listed in Table 1. See the Method section for the description of the model.

\begin{table} [H]
\centering
\begin{tabular}{|c||c|c|}
\hline 
Parameter & Value & Units \\ 
\hline \hline
$\beta_1$ & $4.905\times 10^{-9}$ & s/m \\ 
\hline
$\beta_2$ & $-2.3\times 10^{-26}$ & s$^2$/m \\ 
\hline
$\gamma$ & $1.2\times 10^{-3}$ & W$^{-1}$m$^{-1}$ \\
\hline
$P_\mathrm{sat}$ & $220$ & mW \\
\hline
$\theta$ & ${0.1}$ & - \\
\hline
$\delta_0$ & $0$ & - \\
\hline
$n_\mathrm{sp}$ & $1$ & - \\
\hline
$T_\mathcal{L}$ & $0.71$ & - \\
\hline\hline
$T_\mathcal{G}$ (Fig.3)& $0.377$ & - \\
\hline
$T_\mathcal{G}$ (Fig.4)& $0.44$ & - \\
\hline
$g_0$ (Fig.3)& $1.0$ & - \\
\hline
$g_0$ (Fig.4)& $1.3$ & - \\
\hline
%
$\Delta\nu$ (Fig.4)& 5.6 & THz \\
\hline
$L$ (Fig.3)& $18.23$ & m \\
\hline
$L$ (Fig.4)& 1.823 & m \\
\hline
$\Delta T = t_R/200$ (Fig.3)& 447 & ps \\
\hline
$\Delta T = t_R$ (Fig.4)& 8.94 & ns \\
\hline

\end{tabular} 
\caption{
\textbf{Simulation parameters.} 
$\beta_1$, inverse of the group velocity, 
$\beta_2$, group velocity dispersion,
$\gamma$, fibre nonlinear Kerr coefficient, 
$\theta$ reflection coefficient of the inter-resonator coupler,
$\delta_0$, detuning,
$g_0$, small signal gain,
$P_\mathrm{sat}$, saturation power of the amplifier,
$n_\mathrm{sp}$, spontaneous emission factor,
$T_\mathcal{L}$, transmission passive cavity (without the inter-resonator coupler), 
$T_\mathcal{G}$, cold active cavity transmission (without the inter-resonator coupler),
$\Delta\nu$, width Gaussian spectral filter in the active cavity (FWHM),
$L$, length of the resonators,
$\Delta T$, fast-time simulation window,
$t_R$, round-trip time. We note that for Fig.3, the saturation power was rescaled by a factor $t_R/\Delta T$ to account for a simulation window smaller than the actual round-trip time. 
}
\label{tab:cste}
\end{table}

\section{Coupled equations model}

Under the condition of high effective finesse, the fields propagating in the active and passive cavities can be described by a set of two coupled Ginzburg-Landau equations, which reads\,\cite{xue_super-efficient_2019}:

\begin{align}
    it_R\frac{\partial E_\mathcal{L}}{\partial t} &= \left[\delta_\mathcal{L} - i\alpha + L\frac{\beta_2}{2}\frac{\partial^2}{\partial \tau^2} - \gamma L |E_\mathcal{L}|^2\right] E_\mathcal{L} + \kappa E_\mathcal{G},\\
    it_R\frac{\partial E_\mathcal{G}}{\partial t} &= \left[\delta_\mathcal{G} + ig + L\frac{\beta_2}{2}\frac{\partial^2}{\partial \tau^2} - \gamma L |E_\mathcal{G}|^2\right] E_\mathcal{G} + \kappa E_\mathcal{L}.
\end{align}
In these equations, $t$ is the (slow) time associated with the round-trip evolution
of the electric field envelopes ($E_\mathcal{L,G}$) of the waves propagating in the passive and active cavities.
$\tau$ is a (fast) time variable defined in a co-moving reference frame at the group velocity of the carrier frequency of the fields.
$\alpha$ represents the loss over one round-trip in the uncoupled passive cavity and $g$ the gain in the active one, while $\delta_\mathcal{L}$ and $\delta_\mathcal{G}$ are the phase detunings. The detuning $\delta_0$ is defined as $\delta_0 = \delta_\mathcal{L}-\delta_\mathcal{G}$. Saturation of the gain can be modeled as $g=g_0/(1+\bar{P}/P_\mathrm{sat})$, with $g_0$ the small signal gain, $\bar{P}$ the intracavity average power and $P_\mathrm{sat}$ the saturation power.  

Neglecting the dispersion and the Kerr nonlinearity, this set of equations simplifies to:
\begin{equation}
    i\frac{\partial}{\partial t}\begin{pmatrix}
        E_\mathcal{L}\\
        E_\mathcal{G}
    \end{pmatrix} = \mathcal{\widehat{H}}\begin{pmatrix}
        E_\mathcal{L}\\
        E_\mathcal{G}
    \end{pmatrix} ,
\end{equation}
which is formally equivalent to a Schrödinger equation, with the Hamiltonian:
\begin{equation}
 \mathcal{\widehat{H}} =
\begin{bmatrix}
\delta_\mathcal{L} -i\alpha& \kappa \\
\kappa & \delta_\mathcal{G} +ig
\end{bmatrix}.
\end{equation}

\noindent Under proper gauge transformation, $\mathcal{\widehat{H}}$ transforms into a PT-symmetric Hamiltonian\,\cite{Ozdemir2019}. 
Figs. 1(a)-(b) of the main manuscript show the imaginary and the real parts of the eignevalues $\lambda_{1,2}$ of the $2\times2$ matrix $\mathcal{\widehat{H}}$. The complex frequencies of the corresponding super-modes are given by $\omega_{1,2} = \lambda_{1,2}/t_R$.     

\section{Additional figures}

\subsection{Dynamic scan of the detuning}
\begin{figure}[H]
    \centering
    \includegraphics[width=15cm]{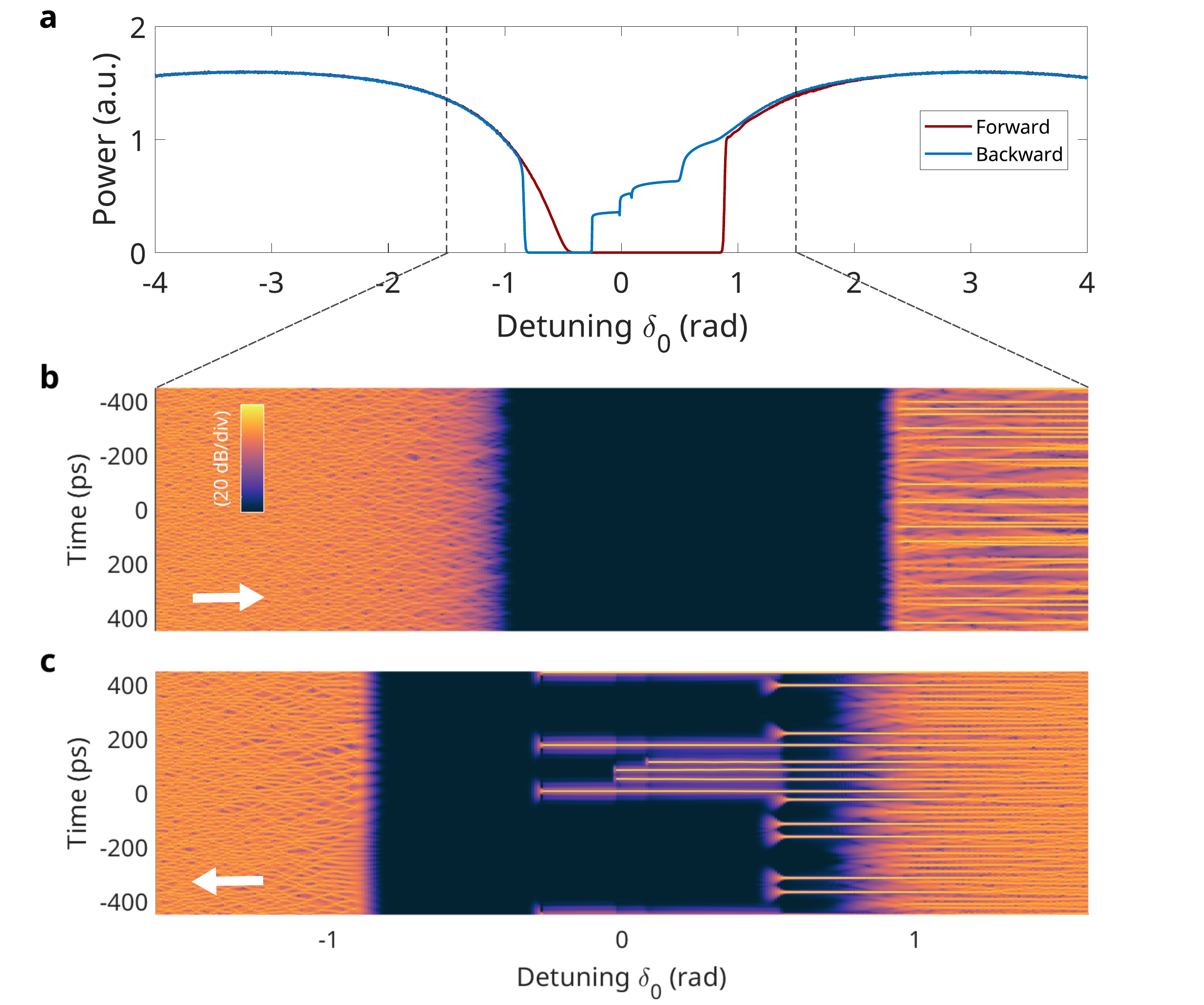}
\caption{Simulation of the intra-cavity dynamics during detuning scans. 
\textbf{a}, Typical average intra-cavity power when the detuning $\delta_0$ is scanned in the forward (red) and backward (blue) directions. The scan rate is $3\pi$\,rad over 50,000 round-trips. However, this rate is too fast to accurately capture the dynamics when the laser turns on, which explains the spurious bistability observed in the negative detuning range, as well as the asymmetry with respect to $\delta_0=0$ in the forward scan curve (not seen in the experiment). Apart from that, the simulations capture well the dynamics observed in the experiment (see Figs\,2.e-f).  
(\textbf{b-c}), Evolution of the intra-cavity power. The white arrows give the direction of the scan. 
In the backward scan, we observe the formation of localized pulses that persist in the central region ($\sim[-0.5, 0.5]$\,rad), which corresponds to the detuning range where lasing does not occur in the linear regime.
This explains the "steps" seen in the corresponding average power curve in the simulation and in the experiments.      
The parameters are $L=18.23$\,m, $\Delta T = t_R/100 = 894$\,ps, $\delta_0=0$, $T_\mathcal{L}=0.71$, $T_\mathcal{G}=0.3763$, $g_0=1.06$ and $P_\mathrm{sat}=220$\,mW. The other parameters are given in Table I.
}
\label{fig:4_scan_dynamics}
\end{figure}

\subsection{Breathing dynamics}
\begin{figure}[H]
    \centering
    \includegraphics[width=\columnwidth]{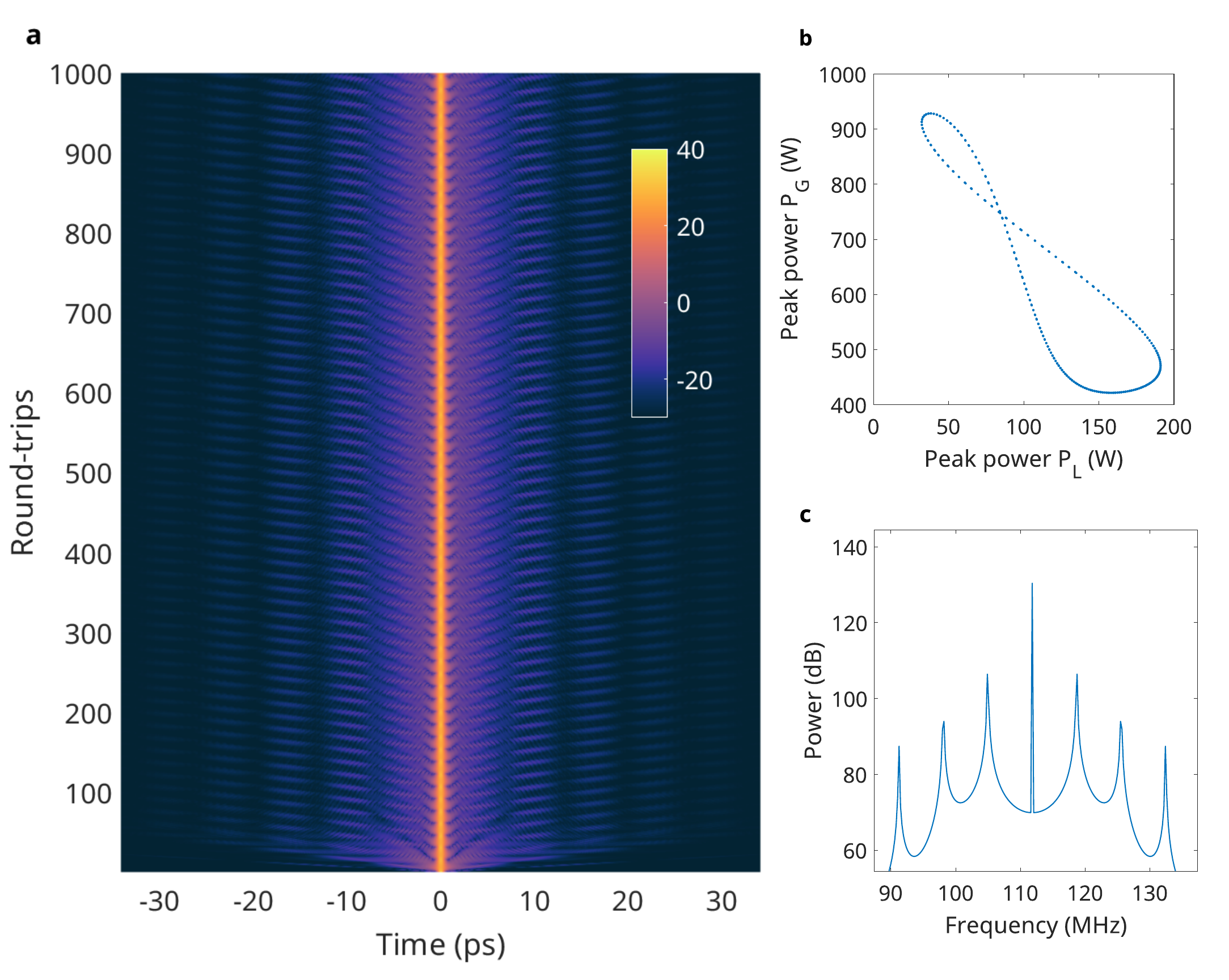}
\caption{Simulation over 1000 round-trips showing breathing dynamics. 
\textbf{a}, Intra-cavity power in the active cavity.
\textbf{b} Limit cycle in the $P_\mathcal{L}-P_\mathcal{G}$ phase plane, plotted for the 500 last round-trips.  
\textbf{c}) Simulated radio-frequency power spectrum at the output of the active cavity (500 last round-trips), around the first beat-note. 
The parameters are $L=1.82$\,m, $\Delta T = t_R=8.9$\,ns, $\delta_0=0$, $T_\mathcal{L}=0.8$, $T_\mathcal{G}=0.6$, $g_0=4$ and $P_\mathrm{sat}=4.2$\,mW. The other parameters are given in Table I. The simulation initial condition is a single pulse in the active cavity.   
}
\label{fig:4_osc_dynamics}
\end{figure}

\subsection{Spontaneous formation for $\delta_0=0.1$}

\begin{figure}[H]
    \centering
    \includegraphics[width=\columnwidth]{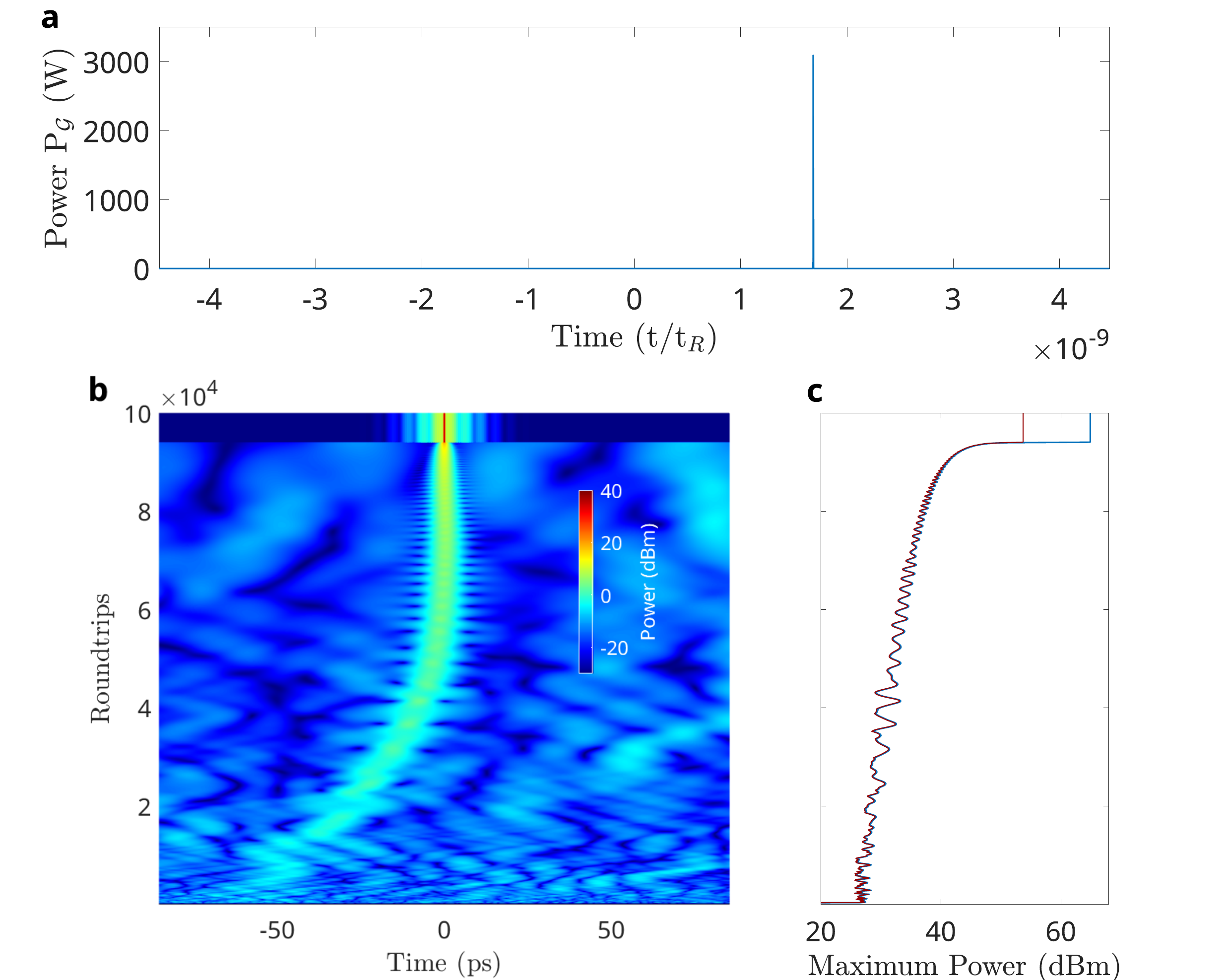}
\caption{Self-starting mode-locking dynamics for $\delta_0=0.1$. 
\textbf{a}, Intra-cavity power after 10$^5$ round-trips
\textbf{b}, Temporal dynamics of spontaneous mode-locking from noise, zoomed around the emerging pulse in the active cavity. 
\textbf{c}, Evolution of the maximum powers in the active (blue) and passive (red) resonators.
The parameters are the same as those used in the simulation shown in Fig. 4 (see Table I).
}
\label{fig:4_starting_dynamics}
\end{figure}

\bibliography{SI_MLPT_biblio}